# Evaluation and statistical correction of area-based heat index forecasts that drive a heatwave warning service.

Nicholas Loveday and Maree Carroll

*Bureau of Meteorology, Melbourne, Victoria, Australia*

*Corresponding author: Nicholas Loveday, nicholas.loveday@bom.gov.au*

**ABSTRACT**

This study evaluates the performance of the area-based, district heatwave forecasts that drive the Australian heatwave warning service. The analysis involves using a recently developed approach of scoring multicategorical forecasts using the FIxed Risk Multicategorical (FIRM) scoring framework. Additionally, we quantify the stability of the district forecasts between forecast updates. Notably, at longer lead times, a discernible overforecast bias exists that leads to issuing severe and extreme heatwave district forecasts too frequently. Consequently, at shorter lead times forecast heatwave categories are frequently downgraded with subsequent revisions. To address these issues, we demonstrate how isotonic regression can be used to conditionally bias correct the district forecasts. Finally, using synthetic experiments, we illustrate that even if an area warning is derived from a perfectly calibrated gridded forecast, the area warning will be biased in most situations. We show how these biases can also be corrected using isotonic regression which could lead to a better warning service. Importantly, the evaluation and bias correction approaches demonstrated in this paper are relevant to forecast parameters other than heat indices.

## 1. Introduction

Heatwaves have devastating impacts on society. They are responsible for increases in detrimental human health issues including renal, cardiovascular, and respiratory problems (Borg et al. 2019; Cheng et al. 2019; Mason et al. 2022) and cause more fatalities than any other natural hazard in Australia (Coates et al. 2022). They also have negative impacts on mental and behavioral disorders (Hansen et al. 2008). Heatwaves impact agriculture (Vitali et al. 2015), infrastructure (McEvoy et al. 2012), and can damage ecological systems (Welbergen et al. 2008; Polazzo et al. 2022). The frequency, intensity, and duration of heatwaves have been observed to be increasing over recent decades globally (Perkins et al. 2012) and these trends are predicted to continue to increase due to global warming (Perkins-Kirkpatrick and Lewis 2020).

Due to the adverse impacts that heatwaves have on Australia, since 2018, the Australian Bureau of Meteorology (hereafter, "the Bureau") has issued maps of categorical heatwave forecasts based on the severity component of the Excess Heat Factor (EHF) (Nairn and Fawcett 2013, 2015) (see Fig. 1 for an example). Starting in 2022, the Bureau began issuing district-based heatwave warnings.

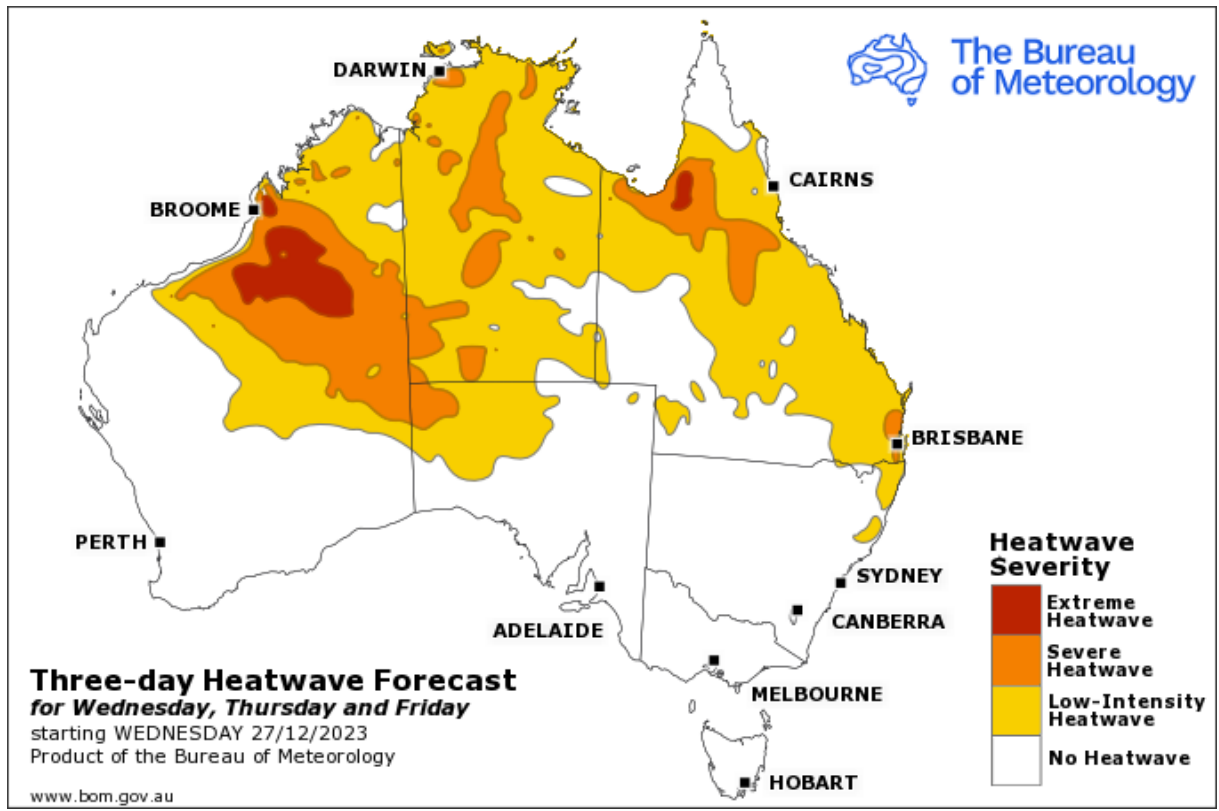

Fig 1. The lead day 2 heatwave forecast for Australia that was issued on 27 December 2023.

The evaluation of heatwave forecasts and warnings is crucial to understand how these high-impact services perform and how they can be improved. In the past, evaluation of these services has taken several forms. Hawkins et al. (2017) surveyed stakeholders to understand the quality of NOAA's National Weather Service (NWS) heat products. Hondula et al. (2022) analyzed how the frequency of heat products issued by the NWS varied spatially across the country. Other research has explored the effectiveness of heat warning services in reducing fatalities and harm (Toloo et al. 2013; Nitschke et al. 2016; Lowe et al. 2016; Heo et al. 2019; Wu et al. 2020).

In contrast to previous work that has sought to evaluate heatwave warning and forecast services, we focus on evaluating the performance of the Bureau's heatwave area-based forecasts against meteorological observations. This is because, in addition to evaluating forecasts and warnings against negative health impacts and customer feedback, it is also important to evaluate the predicted heat index against the observed heat index to drive continuous improvement along the value chain (Hoffmann et al. 2023). Previous research has also focused on evaluating parts of the value chain that occur before the generation of forecast and warning products against meteorological observations, such as coupled forecast models (e.g., Ford et al. 2018). Unlike evaluating forecasts and warnings against negative health impacts, evaluating them against a meteorological index has the advantage that the observed heat index is not influenced by the forecast or warning (e.g., if the warnings are accurate and communicated well, one would expect reduced negative health impacts compared to not having a warning service).

We verify the performance and biases of the categorical warning service using methods that are consistent for a service with a fixed risk threshold, that is, a warning is to be issued when its chance of occurring exceeds a fixed probability threshold. The stability of the warning service between warning revisions is also analyzed. To address performance issues that the analysis highlighted, a post-processing approach is illustrated that conditionally bias corrects the warning service. We show that deriving an area-based warning from a gridded forecast may lead to a biased forecast and we provide a potential solution. Finally, we discuss how using a combination of two different gridded analyses to produce heatwave forecasts is the potential cause for the poorer performance at longer lead times and the sudden change in forecast biases at shorter lead times.

The verification and statistical recalibration approaches as well as the synthetic experiments demonstrated in this paper are not limited to heatwave warnings but could be applied to many area-based warning services derived from gridded forecasts.

## 2. Data

The Excess Heat Factor (EHF) is a heat index based on how high the three-day mean temperature is compared to both the climatology and the preceding 30-days to account for acclimatization. The Bureau's heatwave service uses the EHF severity index ($EHF_{sev}$) for all heatwave forecasts and warnings to quantify the severity of heatwaves (the appendix describes how $EHF_{sev}$ is calculated). Gridded $EHF_{sev}$ heatwave forecasts (see Fig 1 for an example) and corresponding warnings that are derived from the gridded forecast are issued at 0400 UTC (afternoon in Australia) each day during the heatwave season which typically runs from the beginning of October through to the end of March. The grid resolution of both gridded forecast and observed $EHF_{sev}$ data is approximately 5 km. This heat index is for a three-day period and forecasts are issued for seven lead days ($i$ = 0 to 6). The three-day period for lead day $i$, uses daily mean temperature data for days $i-2, i-1, i$. For negative days in the three-day period, gridded Australian Water Availability Project (AWAP) observation data (Jones et al. 2009) are used. The Bureau's official gridded minimum and maximum temperature forecasts are used to derive daily mean temperature forecasts for non-negative lead days. Operational meteorologists are responsible for producing the gridded maximum and minimum temperature forecasts for Australia which are used to produce daily mean temperature forecasts. This means that the lead day 0 forecast comprises of two days of observations and one day of forecast data, and the lead day 1 forecast comprises of one day of observations and two days of forecast data. The Bureau's temperature forecasts aim to predict the mean (i.e., expected) value and are calibrated against a different gridded analysis to AWAP (we discuss the impact of this in Section 7).

Australia is divided into 90 districts. The districts have a mean area of approximately 85343 km$^2$ with a minimum district size of approximately 625 km$^2$ and a maximum district size of approximately 513900 km$^2$. An area-based, district forecast is generated by taking the 0.905 quantile[1] value (spatially) of the gridded forecast values within a district. The district forecast is then converted into one of three categories which is used to drive the warning service:

   i.   Extreme heatwave: $3 \leq EHF_{sev} < \infty$,
   ii.  Severe heatwave: $1 \leq EHF_{sev} < 3$, and
   iii. No heatwave warning: $-\infty < EHF_{sev} < 1$. This includes both the no heatwave and low-intensity heatwave categories that are displayed in Fig 1.

Categorical, district observations are derived in the same way as the district forecasts but using the gridded AWAP observations. The forecast directive for producing district forecasts is to "forecast the highest category for which the probability of observing that category or higher exceeds 50%". Based on the cost-loss model (Thompson 1952; Murphy 1977), this implies that misses and false alarms should be penalized equally (Taggart et al. 2022). A warning is then issued for a district if the district forecast predicts a severe or extreme heatwave during lead days 0 to 3.

We focus our attention on evaluating the area-based, district forecasts that underpin the heatwave warning service. We additionally evaluate the lead day 4 to 6 forecasts as even though they do not currently feed into the warning service, they are made available to key stakeholders as part of a decision support product. We evaluate district forecasts from four heatwave seasons (2020-2021, 2021-2022, 2022-2023, and 2023-2024) retrieved from the Jive verification system (Loveday et al. 2024) using the scores Python package (Leeuwenburg et al. 2024).

## 3. Performance Evaluation

---

[1] The warnings are produced by taking the highest category (inclusive of higher categories) that covers at least 9.5% of the area of the forecast district. This is the equivalent to taking the 0.905 quantile value of the district and converting it to a category.

To evaluate the performance of the categorical heatwave district forecasts that drive the warning service, we compare them against the district $\text{EHF}_{sev}$ observations using the FIxed Risk Multicategory (FIRM) scoring framework (Taggart et al. 2022). The FIRM framework provides a family of scoring functions for multicategorical forecasts, where each scoring function is consistent with a directive based on a fixed risk. A consistent score (Gneiting 2011) is one where the expected score is optimized when a forecaster follows the forecast directive provided to them (e.g., "warn for the highest tornado category for which the probability of observing that category or higher exceeds 10%"). Using a consistent scoring function to evaluate categorical forecasts ensures that producing forecasts based on a specified forecast directive will result in a forecaster producing the same forecast as what they would have forecast to optimize their expected verification score.

To use the FIRM score to evaluate the district forecasts, one must specify:

1. **An increasing sequence of categorical decision thresholds $(\theta_i)_{i=1}^{N}$ for $N+1$ categories.** In the Bureau's heatwave warning service, there are three warning categories, which means there are two $\text{EHF}_{sev}$ categorical decision thresholds. They are $\theta_1 = 1$, which corresponds to the *no warning-severe warning* decision threshold, and $\theta_2 = 3$, which corresponds to the *severe warning-extreme warning* decision threshold.

2. **Threshold weights $(w_i)_{i=1}^{N}$ that correspond to the categorical decision thresholds.** In the Bureau's heatwave warning service, it is more important to correctly decide to warn or not warn than it is to correctly get the forecast district category correct (severe or extreme). For this reason, we assign a weight of 2 for the *no warning-severe warning* decision threshold $w_1$ and a weight of 1 for the *severe warning-extreme warning* decision threshold $w_2$.

3. **The risk threshold $\alpha$, where $0 < \alpha < 1$. The cost of a miss to a false alarm is $\alpha/(1-\alpha)$.** The Bureau has specified that misses and false alarms in the district forecasts should be penalized equally, so the risk threshold is 0.5.

The FIRM scoring function for a single decision threshold $\theta$ is

$$S_{\theta,\alpha}(x,y) = \begin{cases} 1-\alpha, & y < \theta \leq x \quad \text{false alarm} \\ \alpha, & x < \theta \leq y \quad \text{miss} \\ 0 & \text{otherwise} \quad \text{hit or correct negative,} \end{cases} \quad (1)$$

where $x$ is the forecast and $y$ is the observation. For the single decision threshold (binary forecast) case, forecasting an event when $\mathbb{P}(\text{event}) > 1 - \alpha$ will optimize the expected score. With $\alpha$ chosen to be 0.5, the penalty is 0.5 for both false alarms and misses[2].

For multiple decision thresholds, the FIRM scoring function is

$$S(x,y) = \sum_{i=1}^{N} w_i S_{\theta,\alpha}(x,y). \quad (2)$$

Using the specified parameters above, the scoring function can be expressed as a scoring matrix as shown in Table 1. The penalties in the table are symmetric about the diagonal (top left, to bottom right); however, if $\alpha \neq 0.5$, then they would not be symmetric.

|  | Forecast No Heatwave | Forecast Severe | Forecast Extreme |
|---|---|---|---|
| Observed No Heatwave | 0 | 1 | 1.5 |
| Observed Severe | 1 | 0 | 0.5 |
| Observed Extreme | 1.5 | 0.5 | 0 |

Table 1. The FIRM scoring matrix for evaluating district heatwave forecasts. The values indicate the penalty given to each forecast-observation pair. No penalty is given for predicting the correct category.

---

[2] Note that the inequality includes the lower category to align with the heatwave service. This differs from the scoring function in Taggart et al. (2022).

The mean FIRM score is then calculated for the EHF$_{sev}$ district forecasts by taking the mean of the FIRM score across each timestep and district. It is shown for each lead day in Fig. 2. We also create a benchmark forecast which is a hypothetical warning service that never issues a severe or extreme district forecast and calculate its mean FIRM score. By construction in Eq. 1 (which feeds into Eq. 2), the mean FIRM score can be decomposed into under and overforecast components and we display these two components in Fig 2. This decomposition can be visualized on the scoring matrix in Table 1 with overforecast penalties lying to the top right of the diagonal and underforecast penalties lying to the bottom left of the diagonal.

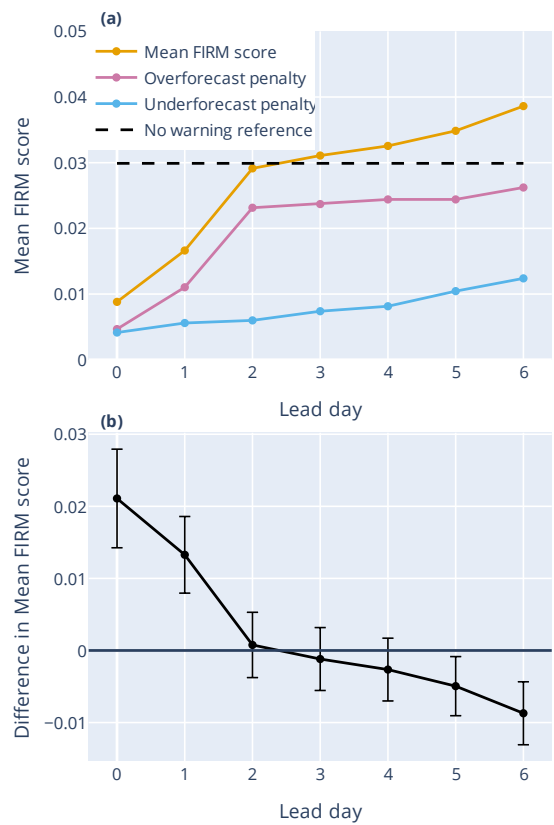

Fig. 2 **a)** The mean FIRM score aggregated across all districts and all times (all four heatwave seasons) is shown for each lead day. Lower scores are better. The overforecast and underforecast components of the mean FIRM score are also displayed. As a benchmark for skill, the black, dashed line shows what the mean FIRM score would have been if the severe or extreme categories were never forecast which we call the "no warning reference". **b)** shows the difference in the mean FIRM score between the district forecasts and the no warning reference with 95% confidence intervals based on the modified Diebold Mariano statistical test (Diebold and Mariano 1995) with the Hering-Genton modification (Hering and Genton 2011) after accounting for spatial dependencies. Positive values indicate that the forecasts performed better than the no warning reference.

We see that nationally, across four warm seasons, the Bureau's district forecasts only outperformed the benchmark at lead days 0-2. The large improvement in performance at lead days 0 and 1 compared to other lead days is unsurprising since those district forecasts contain observation data in their three-day periods. From lead day 1 to lead day 6, the mean FIRM score is dominated by overforecast penalties which implies that there is an overforecast bias in the district forecasts. Possible reasons for these biases will be considered in Section 7 of this paper.

To understand how forecast performance varies spatially, we calculate a skill score for each forecast district defined as

$$\overline{\text{FIRM}}_{ss} = 1 - \frac{\overline{\text{FIRM}}}{\overline{\text{FIRM}}_{\text{ref}}},$$

where $\overline{\text{FIRM}}$ is the mean FIRM score for a district and $\overline{\text{FIRM}}_{\text{ref}}$ is the mean FIRM score for a system that never issues a severe or greater district forecast (i.e., a warning is never issued). We display the skill score for lead days 0 and 2 in Fig. 3.

When $\overline{\text{FIRM}}_{\text{ref}}$ is non-zero, a system that produces perfect district forecasts receives a skill score of 1, while forecast performance that is worse than the reference forecast receives a negative skill score. We also test the null hypothesis that the Bureau's district heatwave forecasts have equal predictive performance to the reference forecast at the 10% level. This is done using the Hering and Genton (2011) modified version of the Diebold-Mariano test statistic (Diebold and Mariano 1995). We use the Diebold-Mariano test statistic since it is suitable when errors are non-Gaussian. Additionally, it accounts for serially and contemporaneously correlated errors.

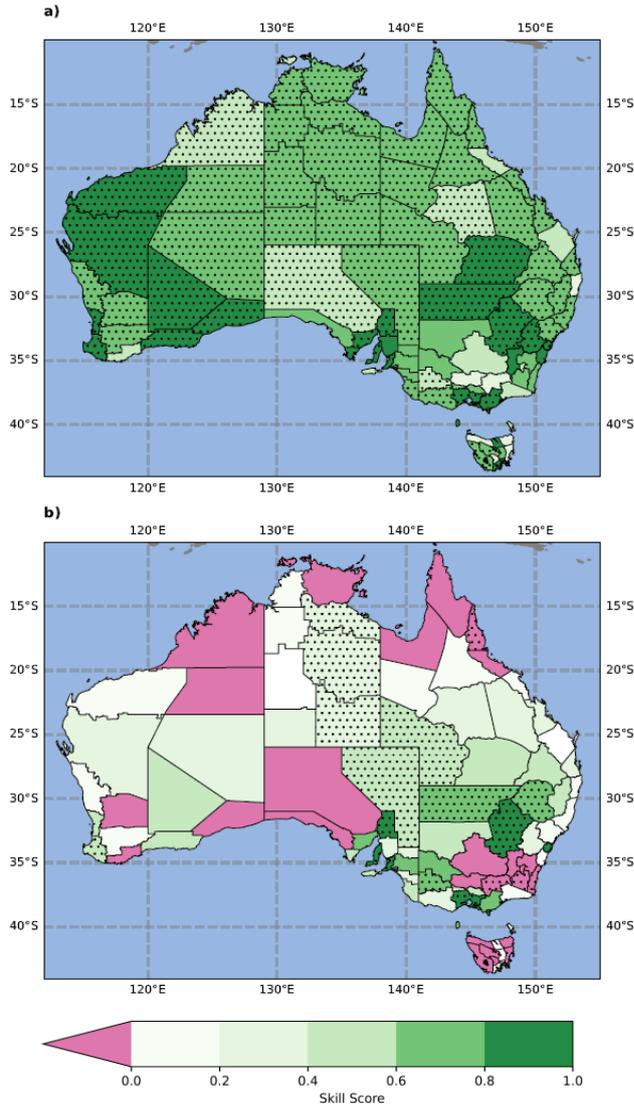

Fig. 3. The FIRM skill score is shown for each forecast district for the four heatwave seasons combined, for **(a)** lead day 0, and **(b)** lead day 2. Green polygons indicate districts that have skill, while pink polygons indicate districts that have no skill compared to the benchmark. The stippling shows where the difference in performance between the district forecasts and the benchmark is statistically significant at the 10% level.

Figure 3a shows that at lead day 0, there is generally skill across most of Australia, with statistically significant differences in predictive performance between the district forecasts and the reference forecast across most of the northern half of Australia. However, by lead day 2, Fig 3b shows that there is no skill over most districts that border the ocean over northern Australia, across Tasmania (the land mass located south of 40°S) and several other districts across mainland Australia. Further, the results show that at lead day 2, we are not confident that there is skill for many forecast districts due to a lack of statistical significance. By lead day 6 (not shown), there were no districts where the forecasts performed better than the benchmark with statistical significance.

## 4. Forecast revision stability of heatwave district forecasts

In the previous section, we showed that forecast performance changed rapidly between lead days 2 and 0 and that this was primarily due to an overforecast bias that increases at longer lead days. Previously, weather forecasters reported that the heatwave district forecasts were "flip-flopping" between forecasts issued across shorter lead times, and often warnings would be canceled during the peak of warm weather. Initially, we quantified this using the flip-flop index (Griffiths et al. 2019) which quantifies how stable the forecasts are with subsequent forecast revisions while not penalizing forecast revisions that only show a trend. The flip-flop index for a forecast sequence $f_1, f_2, \ldots, f_n$ for $n$ forecast issues for a given timestep is defined as,

$$\text{Flip} - \text{Flop Index} = \frac{1}{n-2}\left\{\sum_{i=1}^{n-1}|f_i - f_{i+1}| - (\max_i f_i - \min_i f_i)\right\}$$

where, $i$ is the lead time (e.g., $f_3$ could be a lead day 3 forecast). Lower values indicate more stability (less flip-flopping) between forecast issues.

However, it was found that "flip-flops" were rare and were less frequent at shorter lead times (not shown). Since the flip-flop index does not penalize forecast revisions that only show a trend, instead we calculate the number of forecast district category changes across sequential lead days, as shown in Table 2 and Table 3. This was done regardless of whether it is technically a flip-flop as defined by the flip-flop index.

| Lead day sequence | [6, 5] | [5, 4] | [4, 3] | [3, 2] | [2, 1] | [1, 0] |
|---|---|---|---|---|---|---|
| **Category decrease count** | 643 | 437 | 444 | 336 | 1027 | 558 |
| **Category increase count** | 655 | 612 | 444 | 394 | 173 | 209 |
| **Category change count** | 1298 | 1049 | 1049 | 517 | 1200 | 767 |
| **Ratio (category decrease/increase)** | 0.98 | 0.71 | 1.0 | 0.85 | 5.94 | 2.67 |
| **Category no change count** | 1977 | 2158 | 2318 | 2450 | 1849 | 1641 |

Table 2. Counts for category increases, decreases, and total changes across the four heatwave seasons for all districts are shown for district forecast revisions across sequential lead days, $[i, i-1]$. Additionally, the ratio of the decreases to increases is shown. The "category" counts include changes between the three categories *no heatwave warning*, *severe heatwave*, and *extreme heatwave*. The Category no change count shows the number of times that the category did not change between forecast revisions when *severe* or greater was forecast.

| Lead day sequence | [6, 5] | [5, 4] | [4, 3] | [3, 2] | [2, 1] | [1, 0] |
|---|---|---|---|---|---|---|
| **Heatwave cancellation count** | 568 | 375 | 368 | 291 | 829 | 480 |
| **Heatwave upgrade count** | 573 | 500 | 374 | 342 | 160 | 188 |
| **Heatwave change count** | 1141 | 875 | 742 | 633 | 989 | 688 |
| **Ratio (heatwave cancellation/update)** | 0.99 | 0.75 | 0.98 | 0.85 | 5.18 | 2.55 |
| **Heatwave no change count** | 2134 | 2332 | 2464 | 2547 | 2060 | 1740 |

Table 3. As per Table 2, but where counts only include changes between *no heatwave warning* and a severe or greater heatwave. The heatwave change count is the sum of the heatwave cancellation and upgrade counts.

The total number of changes in district forecast categories was generally higher for sequential district forecast revisions at longer lead days than shorter lead days. The exception is the number of revisions that occurred between the lead day 2 and lead day 1 district forecasts, where we see a spike in the number of category changes. This spike is attributed to a large jump in district forecast category decreases. This behavior occurs both when looking across the three categories (Table 2) and when we split it into two categories; *heatwave* (which includes *severe* and *extreme*) and *no heatwave warning* (Table 3). When all the cases where there was a heatwave forecast at lead day 2, 42% of the time by lead day 0, there was no heatwave forecast (not shown).

While meteorologists suggested that qualitatively, there was significant flip-flopping at shorter lead days, the flip-flop index did not indicate this. Instead, these results show that district forecast categories are frequently being downgraded at shorter lead days, far more frequently than they are upgraded. These downgrades are responsible for warnings being canceled, often during the peak of the warm weather. This could lead to decreased trust in the heatwave warning service and may have implications for how people respond to the warnings.

## 5. Statistically correcting the district-based heatwave forecasts

To understand the nature of the biases of the $EHF_{sev}$ district forecast across Australia, we perform isotonic regression to minimize the absolute loss (Fig 4). This is consistent with the service and how we have set up the FIRM score with an $\alpha$ of 0.5. Isotonic regression is a free-form curve that minimizes the loss between forecasts and observations, subject to the constraint that the curve is non-decreasing. It has an advantage over many other regression techniques since it can recalibrate non-linear, conditionally biased forecasts towards a chosen functional (e.g., 50$^{th}$ percentile in this case) (Jordan et al. 2022).

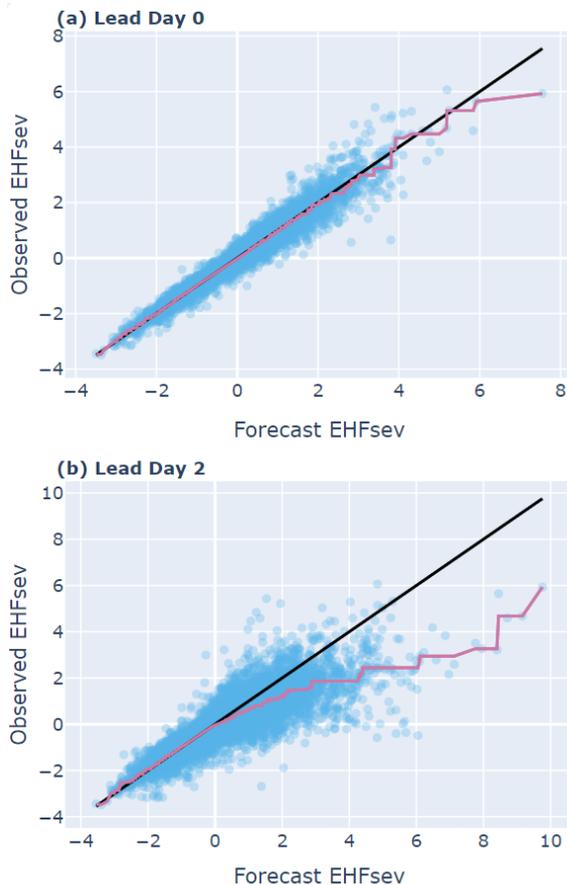

Fig. 4. Forecast $EHF_{sev}$ is displayed against observed $EHF_{sev}$ as blue dots for all districts and all four heatwave seasons for **(a)** lead day 0, and **(b)** lead day 2. The solid pink line is the isotonic regression line fitted to target the median of the underlying data.

We can see that at lead day 0, there is almost no bias, but at lead day 2, a conditional bias begins to emerge when the $EHF_{sev}$ district forecasts are positive (Fig 4). This is consistent with the overforecast penalty results highlighted by the mean FIRM score in Fig 2, and the spike in forecast district category decreases that occur at shorter lead days noted in Table 2.

Next, we demonstrate how isotonic regression can be used to conditionally bias correct the $EHF_{sev}$ district forecasts. First, we split our data into training (2020-2021, 2021-2022, 2022-2023 heatwave seasons) and test (2023-2024 heatwave season) datasets. Appendix B gives counts of the number of severe or higher and extreme observations and lead day 2 district forecasts in the training and test datasets. Then at each lead day and each forecast district in our training dataset, we perform isotonic regression targeting the median. The conditional biases vary across different districts and slightly across lead days (not shown). Since isotonic regression cannot recalibrate a forecast that has a value outside the range of the training dataset, we perform linear regression on positive values of the isotonic regression curve and use this function so that we can recalibrate $EHF_{sev}$ district forecasts that are higher than our training dataset. It is possible for the recalibrated district forecast value to fall outside the range of the gridded forecast values for that district which could pose communication challenges. To address this, we add a constraint so that the district forecast can only be recalibrated up to the maximum value or down to the minimum value in the gridded forecast for the district. In the 2023-2024 heatwave season, this constraint was enforced between 7 and 30 times depending on the lead day. This corresponds to less than 0.2% of all district forecasts produced.

We now apply our recalibration function to the 2023-2024 heatwave season $EHF_{sev}$ district forecasts. Figure 5a compares the mean FIRM score for the recalibrated and the raw district forecasts. The calibrated district forecasts are shown to have skill at all lead days, except for lead day 6, and perform better than the raw district forecasts for lead days 1-6. At lead day 6, calibrated district forecasts outperformed the raw lead day 2 district forecasts. We also calculate the difference in mean FIRM score between the raw and calibrated district forecasts (Fig 5b). 95% confidence bands are calculated using the Diebold Mariano test statistic after the mean in the score differences is taken at each date to account for any spatial dependency. The performance improvements were statistically significant for all lead days, except for lead days 0 and 1.

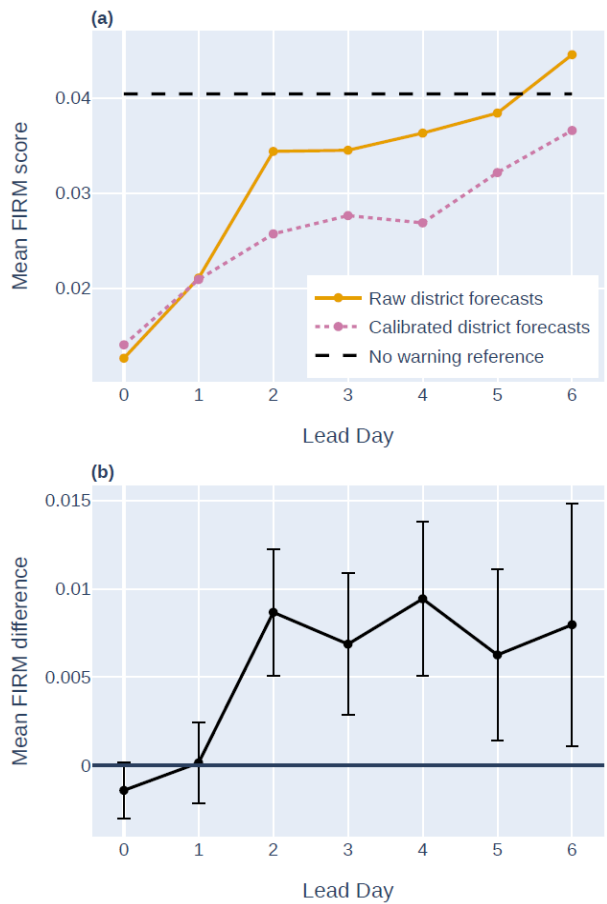

Fig. 5. **a)** Mean FIRM scores aggregated across all forecast districts for the 2023-2024 heatwave season for each lead day. The mean FIRM score for raw district forecasts is shown by the solid orange line while the mean FIRM score for the recalibrated district forecast is shown by the dotted pink line. The black, dashed line is the mean FIRM score of a reference forecast district system that never forecasts *severe* or *extreme*. **b)** The difference in the mean FIRM score between the raw district forecasts and the calibrated district forecasts with 95% error bars based on the Diebold Mariano test statistic with the Hering-Genton modification. Positive values indicate that the calibrated district forecasts performed better than the raw forecasts.

One limitation of this approach is that the conditional bias correction function may not be optimal if forecast systems or gridded analysis products change in the future. A reforecast of the forecast system of suitable length would need to be generated to derive a new conditional bias correction function. However, one advantage of this approach is that if in the future, the risk threshold $\alpha$ of the district forecast service needs to be modified, this conditional bias correction approach can easily be run again to target the relevant quantile that is consistent with the chosen $\alpha$.

## 6. Challenges with deriving area warnings from gridded forecasts

Ideally, the underlying gridded $EHF_{sev}$ forecast would be recalibrated across Australia (i.e., the gridded forecast in Fig 1), and the area-based, district forecasts could be derived from those. However, even if the underlying grid was perfectly (re)calibrated and we had many forecast-observation pairs, in this section we show that an area-based warning derived from a perfectly calibrated gridded forecast can still result in a biased warning. In addition to area-based heatwave warnings, the Bureau issues other warning products that are also derived from gridded forecasts using a similar process, including fire weather and marine wind warnings, that may experience a similar problem. Using a synthetic experiment, we illustrate that deriving a warning for an area from an *ideal* gridded forecast that is perfectly calibrated, may lead to a biased warning service. The *ideal* predictive distribution is one that is indistinguishable from the distribution that the observations are drawn from (Gneiting et al. 2007). To generate the *ideal* single-valued forecast for a percentile, we take a specified percentile value from the *ideal* predictive distribution that we use as our *ideal* single-valued forecast.

We create synthetic observations in the following way.

1. Suppose we have a hypothetical one-dimensional[3] forecast district with 400 grid points $z$ that run from south to north. We simulate a climatological north-south temperature gradient in $EHF_{sev}$ in the southern hemisphere with a linear gradient such that $C(z) = 0.01z$, where $0 \leq z \leq 399$. In our synthetic experiment, we have 10,000 timesteps $t$. The climatological $EHF_{sev}$ value at point $z$ and time $t$ is then $E_{clim}(z,t) = C(z)$ since $C(z)$ is identical for all timesteps.

2. We then simulate the influence of day-to-day changes on the district with a synoptic forcing at each timestep with the $EHF_{sev}$ value at point $z$ and time $t$ being $E_{synop}(z,t) = C(z)U_t$, where $U_t \sim \text{Unif}[0,1]$.

3. Then, on each day we simulate the impact of random local temperature fluctuations (e.g., from convective thunderstorms) where the observed $EHF_{sev}$ $E(z,t)$ is a random value drawn from the normal distribution $\mathcal{N}(E_{synop}(z,t), 1)$ with the mean equal to $E_{synop}(z,t)$ at that point, and a standard deviation of 1.

These three steps are graphically illustrated in Fig 6 for the first 10 timesteps.

---

[3] In practice a forecast district will be a 2-dimensional area, but we reduce it to 1-dimension to simplify the experiment.

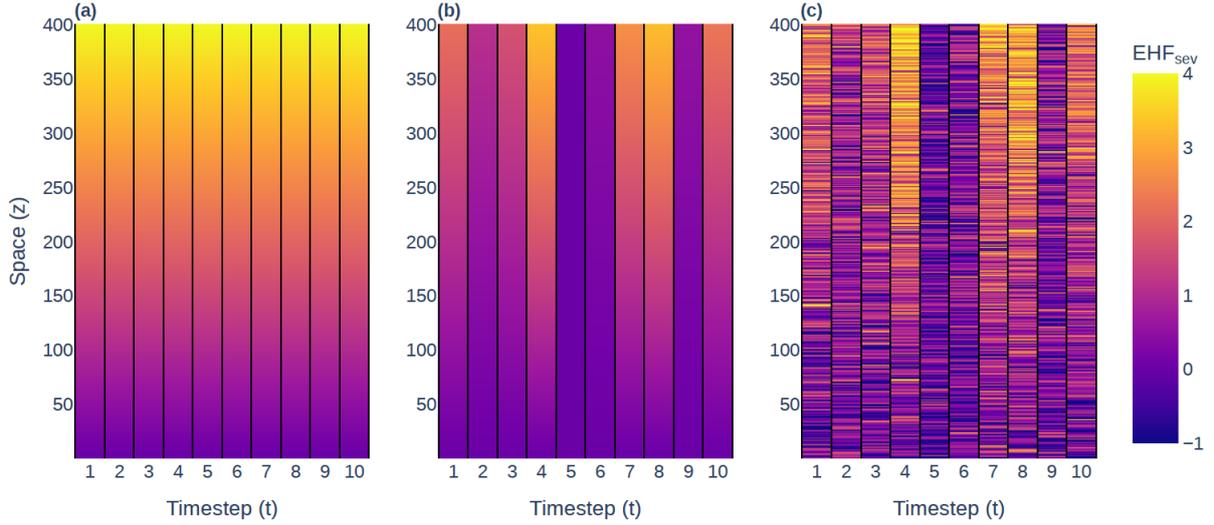

Fig. 6. A graphical representation of the three steps to generate the synthetic observations. The EHF$_{sev}$ values for the 1-dimensional forecast district are shown for the first 10 timesteps only. a) step 1: the background temperature gradient, b) step 2: the influence of a synoptic forcing, and c) step 3: the final observations that are drawn from a normal distribution that are centered on the values in b).

Suppose that a forecaster has perfect knowledge of the climatological background state, the synoptic forcing, and the scale of local temperature fluctuations. For each time $t$ and location $z$, this forecaster issues the ideal probabilistic forecast $F_{z,t} = \mathcal{N}(E_{\text{synop}}(z,t), 1)$ with respect to this information. That is, the forecaster's predictive distribution is identical to the distribution from which observations are drawn. Since we used a normal distribution in step 3 above, to produce an ideal, single-valued 50[th] percentile forecast $G_{50}$ for each point (in contrast to a predictive distribution at each point), the forecaster predicts $E_{synop}(z,t)$.

The synthetic gridded forecast $G_{50}$ is consistent for forecasting the median of the predictive distribution at each grid point. To create synthetic district forecasts $A_{0.905}$, we take the 0.905 quantile value that occurs spatially in each district for each timestep of the synthetic forecasts $G_{50}$. Similarly, we create synthetic district observations $Y_{0.905}$, by taking the 0.905 quantile value that occurs spatially at each timestep in the synthetic gridded observations $E$. Fig 7a shows a time series of the synthetic district forecasts $A_{0.905}$ and the synthetic district observations $Y_{0.905}$ for the first 70 days.

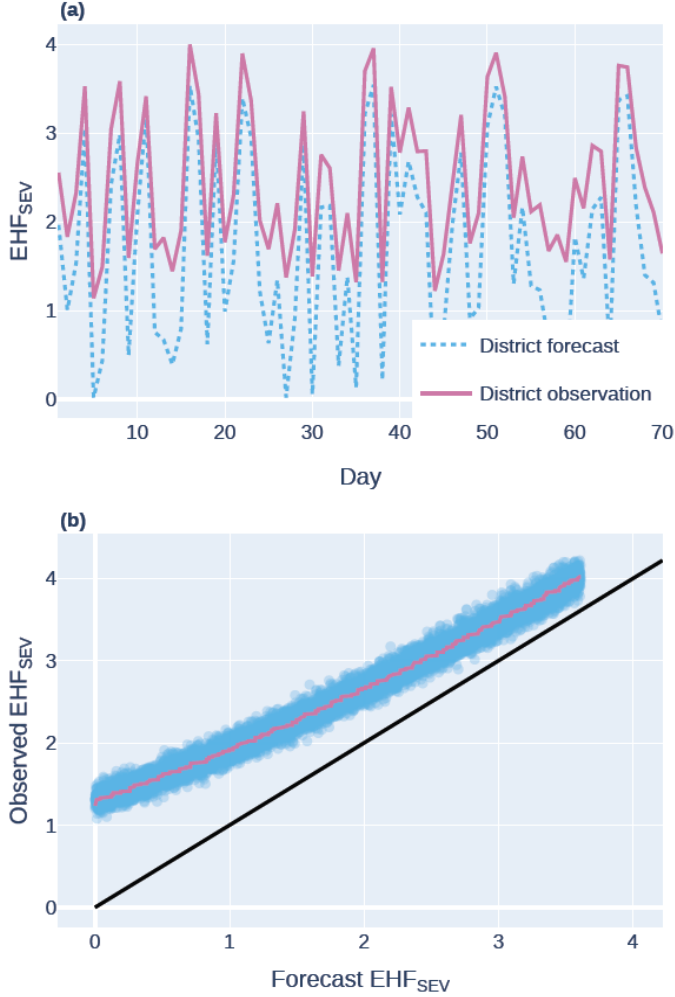

Fig. 7. a) Time series of synthetic EHF$_{sev}$ district forecasts $A_{0.905}$ (dotted blue) and synthetic EHF$_{sev}$ district observations $Y_{0.905}$ (solid pink) across the first 70 days of the experiment. b) Synthetic EHF$_{sev}$ district forecasts are displayed against synthetic EHF$_{sev}$ district observations for 10,000 dates (blue markers). Isotonic regression that minimizes the absolute loss (targeting the median functional) is shown by the pink line.

Figure 7 shows that despite having perfectly calibrated (relative to the available information set), single-valued, gridded forecasts $G_{50}$, when we derive area-based forecasts $A_{0.905}$, the area-based forecasts and warnings are biased compared to area-based observations $Y_{0.905}$. In our simple synthetic experiment, Fig 7b shows that there is a non-linear underforecast bias. The percentile to derive the district forecast or warning influences the bias. For example, if the area-based forecast is derived from the 0.1 quantile value that occurs spatially in $G_{50}$ at each timestep (i.e., $A_{0.1}$), rather than the 0.905 quantile value, then there will be an overforecast bias instead. Additionally, if the standard deviation in the random temperature fluctuations is increased, forecast sharpness in $F_{z,t}$ is reduced and the biases in the district forecasts $A_{0.905}$ increase (not shown). Note that while we kept the standard deviation constant in this experiment for simplicity, for real EHF$_{sev}$ forecasts, forecast sharpness would likely decrease for more extreme EHF$_{sev}$ forecasts since EHF has a quadratic response to daily temperature inputs (see the Appendix for how EHF is derived). Additionally, in the real world, forecast sharpness will likely vary with each day.

To show how isotonic regression can be used to address this bias problem, we take the first 100 days of our synthetic datasets as our training period. We perform isotonic regression on our training dataset to derive a regression function to recalibrate future district forecasts. We then apply this regression function to our test dataset which corresponds to the following 9,900 days of data. For simplicity, since we will only evaluate the categorical performance, forecast values are clipped to have a minimum value of -1 or a maximum value of 3.5 if the district forecast value in the test forecast dataset is outside the range of the training dataset. We then evaluate the performance of the synthetic district forecasts in the test dataset against the recalibrated forecasts. We do this by performing bootstrapping, where we take 1,000 samples, each with 1,000 timesteps, and calculate the mean FIRM score using the setup as described in Section 3 and derive 95% confidence intervals. The mean FIRM score (with lower and upper confidence intervals) for the raw synthetic district forecast is 0.358 (0.337, 0.380), and for the recalibrated district forecast is 0.019 (0.013, 0.025). This indicates a large improvement is gained by recalibrating the synthetic district forecasts. This synthetic experiment highlights the potential need to recalibrate area-based forecasts and warnings that are derived from gridded forecasts, even when the gridded forecasts are perfectly calibrated. By adding a recalibration step when creating the area-based forecasts and warnings we can prevent the dilemma forecasters face in deciding if they should optimize the accuracy of the forecasts or rather produce forecasts that optimize the accuracy of the warnings. This approach raises the chance of a potential communication challenge since it is theoretically possible for the warning category to fall outside the range of values forecast for the district if the forecast uncertainty is very high and if the accuracy of both forecasts and warnings are both being independently optimized. However, a consistency constraint like in section 5 could be considered to address this potential issue.

## 7. Discussion of the causes of bias and future research opportunities

While we have focused on evaluating and post-processing the district forecasts, the cause of the biases impacting the underlying heatwave forecast grids should be investigated in the future. This section discusses potential causes of the biases.

The official temperature forecasts that are used to derive the $EHF_{sev}$ index are generated from the Gridded Operational Consensus Forecast (GOCF) (Engel and Ebert 2012) which is a bias-corrected ensemble of multiple NWP inputs. Additional adjustments are often made by weather forecasters to the gridded forecasts where they believe they can improve the accuracy (Just and Foley 2020). One initial problem that arises is that GOCF is calibrated against the Mesoscale Surface Analysis System (MSAS) (Glowacki et al. 2012), but AWAP is used for observations in the heatwave service. At the time when GOCF was developed, MSAS provided a higher resolution gridded analyses, had other parameters (e.g., wind speed), and had hourly observations. In contrast, AWAP was used when developing the $EHF_{sev}$ index as it uses long term climatological data (see eq. A5 in the Appendix) which the AWAP data provides. This means that forecasts calibrated against MSAS may appear biased when compared to AWAP as they use very different methods to produce their analyses, including the treatment of topography (Fawcett and Hume 2010). Additionally, when weather forecasters adjust the gridded forecasts, they are left with the dilemma as to if they should optimize the temperature forecasts which are evaluated against weather stations or if they should adjust the gridded temperature forecasts to be optimized for the heatwave forecasts.

Additionally, the temperature forecast grids are regridded to have the same grid spacing as the AWAP grids before the $EHF_{sev}$ forecast grids are created. This regridding affects the forecasts, particularly around topographical features such as mountainous areas and coastlines. To address these differences, the heatwave forecast service does a simple bias correction based on the difference in the daily mean temperature between AWAP and the official forecasts over the 2015-2016 two-year period. In addition to the bias correction grid affecting coastlines and mountainous areas, it affects areas where observational uncertainty is high due to the sparseness of the weather station network. However, there are two limitations to this approach. First, the spatial resolution of the GOCF guidance doubled in August 2020, but the bias correction grid was not updated. Secondly, Fig 4b shows that the district forecasts are conditionally biased, particularly for larger $EHF_{sev}$ values, and thus a simple bias correction is not enough to correct the conditional biases.

This research raises several research questions.

1. What is the best method to conditionally bias correct the gridded heatwave forecasts (i.e., Fig 1)? While we have used an isotonic regression approach to correct the district forecast, there is also the opportunity to explore neural network image processing techniques for improving the gridded forecasts.

2. What is the quality of the gridded analyses for analyzing temperature extremes? This is important to understand since gridded analyses are used as part of the heatwave service, but also so that we know how much to trust the verification results.

3. If MSAS (or its future replacement) is suitable for analyzing temperature extremes, could it be used for the observations in the lead day 0 and 1 grids instead of AWAP? Noting that it would need a statistical adjustment to capture the true diurnal range since it only has an hourly temporal resolution. Additionally, the $T_{95}$ (see Appendix) grid may need adjusting to be consistent with the MSAS data. If this was done, then MSAS could also be used as observations for verification purposes.

4. Based on health data, is there a more acceptable risk threshold $(1 - \alpha)$ at which to issue a warning to minimize loss of life and reduce negative health impacts, and what impact is the current heatwave warning having on reducing heat related illnesses and fatalities?

While there are uncertainties in the quality of the AWAP observations that we used as the truth in the verification results that we presented, our post-processing bias correction approach increased the consistency of the district forecasts across lead days. Even if the AWAP observations are biased compared to reality, we suspect that a biased forecast that is consistent across lead days may be more useful for most users compared to one that is inconsistent across lead days, particularly if users can self-calibrate what impacts are associated with a particular forecast. We leave this as an open research question for social scientists to address.

## 8. Conclusion

Since heatwaves pose a significant threat to both the Australian and international communities and are expected to worsen in the future, we must improve heatwave warning services. This includes improving gridded observation datasets, calibrating warning services, evaluating the performance of forecasts and warnings against health data, and for social scientists to continue investigating the best ways to construct and communicate effective warnings.

The categorical heatwave district forecasts used to produce the Bureau of Meteorology's heatwave warning service were analyzed to understand their quality using the FIRM score. The results showed that there is an overforecast bias at longer lead days which reduced forecast performance significantly. The overforecast bias is likely due to the use of different gridded analyses being used to generate and evaluate the district forecasts. More research can be done in the future to quantify the impact of using different grids, resolve these tensions, and improve the heatwave forecasts.

The stability of the district forecasts was explored, and we found that at short lead days, the heatwave district categories were being downgraded far more frequently than they were being upgraded. We reduced the biases and improved the stability of the district forecasts using a post-processing bias correction technique that significantly improved the performance of the district forecasts across lead days 1 to 6.

We demonstrated that even if gridded forecasts are perfectly calibrated, deriving an area warning from gridded forecasts can lead to a biased warning service, especially when a quantile far from the median is used to derive the warning. We showed that isotonic regression can correct this issue and is suitable when a sufficiently long calibration dataset is available.

Finally, this paper demonstrated how area-based multicategorical forecasts and warnings can be evaluated objectively using consistent scores when a risk threshold is provided to forecasters and/or forecast system developers. While area-based forecasts and warnings are a useful communication tool, we have highlighted challenges with deriving area-based forecasts and warnings from gridded forecasts and showed that care must be taken when designing these services.

*Acknowledgments.*

We thank Monica Long, Cassandra Rogers, Robert Taggart, Deryn Griffiths, and Beth Ebert from the Bureau of Meteorology for their constructive comments on earlier versions of this manuscript. We would also like to thank Paul Iñiguez and two anonymous reviewers who provided useful feedback.

*Data Availability Statement.*

All code and data for this analysis are available at https://github.com/nicholasloveday/heatwave_warnings. Code for the FIRM score, Diebold Mariano test statistic, flip-flop index, and isotonic regression can be found at https://github.com/nci/scores.

# APPENDIX A

## Calculation of excess heat factor severity (EHF$_{sev}$)

EHF$_{sev}$ is calculated as follows. Significant excess heat (EHIsig) is the three-day mean temperature anomaly relative to the 95th percentile climatology value at each location. The daily mean temperature ($T$) is calculated as the average of the maximum and minimum temperatures recorded across the 24-hour period (9 am to 9 am local time) for each grid cell where the maximum temperature usually precedes the minimum temperature (°C). The significant excess heat index at a given grid cell for a particular date is calculated as

$$\text{EHI}_{\text{SIG}} = \frac{T_{i-2}+T_{i-1}+T_i}{3} - T_{95} \tag{A1}$$

where $i$ is the day in the three-day heatwave period. $T_{95}$ is the 95th percentile of daily mean temperature $T_i$ for a grid cell for the climate reference period 1971-2000 across all days of the year.

Heat stress (acclimatization) ($\text{EHI}_{\text{accl}}$) for a given grid cell on a particular date is the three-day mean temperature anomaly relative to the previous 30-day mean temperature value. This is calculated as:

$$\text{EHI}_{\text{accl}} = \frac{T_{i-2}+T_{i-1}+T_i}{3} - \frac{T_{i-3}+T_{i-4}+\cdots+T_{i-33}}{30} \tag{A2}$$

where $i = 1$ is the last day in the three-day heatwave period.

Excess heat factor (EHF) is the combined effect of EHIsig and EHIaccl and is calculated as

$$\text{EHF} = \text{EHI}_{\text{sig}} \times \max(1, \text{EHI}_{accl}). \tag{A3}$$

Heatwave conditions exist when EHF is positive. Note that Nairn and Fawcett (2015) use the equation

$$\text{EHF} = \max(0, \text{EHI}_{\text{sig}}) \times \max(1, \text{EHI}_{\text{accl}}), \tag{A4}$$

but the actual forecast service uses the initial EHF equation (Equation A3).

The EHF threshold for a severe heatwave varies by location and is taken to be the 85th percentile ($\text{EHF}_{85}$) at each grid cell for all positive EHF values during the reference period 1958-2016. To categorize heatwave severity, i.e. $\text{EHF}_{\text{sev}}$, EHF is compared to the 85th percentile value at each grid cell as

$$\text{EHF}_{\text{sev}} = \frac{\text{EHF}}{\text{EHF}_{85}}. \tag{A5}$$

# APPENDIX B

## Severity counts in training and testing data for post processing

Table A1 shows the number of extreme, and severe or higher forecasts and observations that were in the training and test datasets. Only 6 out of the 90 districts did not have a severe or higher heatwave event in the observation dataset and there was only one district that did not have a lead day 2 severe heatwave forecast (King Island). In contrast, only 42 out of 90 districts had an extreme category forecast at lead day 2. Out of the 90 districts, 25 had a higher maximum EHFsev value in the 2023-2024 lead day 2 forecast test dataset than the highest district forecast during the three preceding seasons in the test dataset. Out of these 25 districts, 9 of them had a higher category in the test dataset than the train dataset.

|  | **Observation** | **Forecast (lead day 2)** |
|---|---|---|
| **Severe or higher counts in training dataset** | 1246 | 1918 |
| **Severe or higher counts in test dataset** | 86 | 278 |
| **Extreme counts in training dataset** | 645 | 971 |
| **Extreme counts in test dataset** | 34 | 84 |

Table A1. The counts of extreme, and severe or higher district forecasts and observations. The counts are shown for the train and test datasets. Only the counts for the lead day 2 district forecasts are shown.